\documentclass[prd,amsmath,twocolumn]{revtex4}
\usepackage{graphicx}
\usepackage{color}
\usepackage{amsfonts}
\usepackage{bm}
\usepackage{natbib}
\newcommand{\be}{\begin{equation}}
\newcommand{\ee}{\end{equation}}
\newcommand{\ba}{\begin{eqnarray}}
\newcommand{\ea}{\end{eqnarray}}
\newcommand{\nn}{\nonumber}

\def\vecp{\bm{p}}
\def\vecb{\bm{b}}
\def\veca{\bm{a}}
\def\veck{\bm{k}}

\newcommand{\gsim}{\raise.3ex\hbox{$>$\kern-.75em\lower1ex\hbox{$\sim$}}}
\newcommand{\lsim}{\raise.3ex\hbox{$<$\kern-.75em\lower1ex\hbox{$\sim$}}}

\renewcommand{\vec}[1]{\boldsymbol{#1}}

\begin{document}
\title{
Light from Cosmic Strings
}

\author{Dani\`ele A.~Steer}
\affiliation{APC\footnote{Universit\'e 
Paris-Diderot, CNRS/IN2P3,  
CEA/IRFU and Observatoire de Paris},10 rue Alice Domon et L\'eonie Duquet,
 75205 Paris Cedex 13, France}
\author{Tanmay Vachaspati}
\affiliation{Physics Department, Arizona State University, Tempe, AZ 85287, USA}

\begin{abstract}
The time-dependent metric of a cosmic string leads to an effective 
interaction between the string and photons -- the ``gravitational 
Aharonov-Bohm'' effect -- and causes cosmic strings to emit light. 
We evaluate the radiation of pairs of photons from cosmic strings 
and find that the emission from cusps, kinks and kink-kink collisions 
occurs with a flat spectrum at all frequencies up to the string scale. 
Further, cusps emit a beam of photons, kinks emit along a curve, and 
the emission at a kink-kink collision is in all directions. 
The emission of light from cosmic strings could provide an important 
new observational signature of cosmic strings that is within reach 
of current experiments for a range of string tensions.
\end{abstract}

\maketitle

\section{Introduction}
\label{sec:introduction}

Cosmic strings are possible remnants from the early universe
(for a review see \cite{VilenkinShellard}) and there is significant
effort to try and detect them. A positive detection of cosmic
strings will open up a window to very high energy fundamental
physics and can potentially have strong implications for
astrophysical processes. Hence it is of great interest 
to continue
to discover new observational signatures of cosmic strings,
as well as to refine features of known signatures.  In this paper 
we address the radiation of photons by cosmic strings.

There is an extensive literature on gravitational radiation from 
cosmic strings, particularly motivated by upcoming and future 
gravitational wave detectors. More relevant to the work presented 
here, however, is the analysis in \cite{JonesSmith:2009ti} and 
\cite{Chu:2010zzb} of the emission of particles due to the 
time-dependent metric of cosmic strings from the viewpoint of 
Aharonov-Bohm radiation. The case of photon emission -- which we 
treat in the present paper -- was not explicitly discussed there.  
A crucial feature which
emerged in these calculations is that cusps and kinks on cosmic
strings emit radiation with a flat spectrum all the way up
to the string scale. However, those results were based on 
studying two rather specific loop configurations with cusps and kinks.
As we show here in more generality (namely for any loop configuration) 
the flat spectrum also applies to the emission
of photons from cusps and kinks, as well as kink-kink collisions. 
Thus light emitted from cosmic strings in this way leads to a new and 
observable signature of cosmic strings that is completely independent 
of the details of the underlying particle physics model. As we shall 
see, the effect is small, however, being proportional to
$(G\mu)^2$ where $G$ is Newton's constant and $\mu$ the string
tension. Despite that, since photons are being emitted, it may be more 
easily measurable than, say, the gravitational wave (GW) bursts also 
emitted by cusps and kinks. 

The total power emitted in scalar particles from cosmic strings due 
to their gravitational coupling was first considered in \cite{Garriga:1989bx}, 
using formalism developed in \cite{Frieman:1985fr}. 
In this paper we calculate the differential power emitted in photons
from cosmic strings due to the gravitational coupling. We call
this the ``gravitational Aharonov-Bohm'' effect because the
metric is flat everywhere except at the location of the string,
and is closely analogous to the case of the electromagnetic
Aharonov-Bohm effect due to a thin solenoid. In Sec.~\ref{sec:gAB}
we set up the calculation and evaluate the invariant matrix
element for the production of two photons. The emission
is dominant in three cases -- from cusps, kinks and kink-kink 
collisions. Integrals relevant to these cases are evaluated in 
Sec.~\ref{sec:integrals}. In Sec.~\ref{power} we find the power 
emitted from cusps, kinks and kink-kink collisions on strings. 
Our results are summarized in Sec.~\ref{conclusions}, where we 
also consider observational signatures. Our numerical estimate 
in Eq.~(\ref{calN}) indicates that light from cosmic strings may 
potentially be detectable by current detectors for a range of 
string tensions.

\section{Gravitational Aharonov-Bohm}
\label{sec:gAB}

The gravitational field of a cosmic string is characterized
by the parameter $G\mu$ which is constrained to be less than
$\sim 10^{-7}$. Hence it is sufficient to consider the case of a
weak gravitational field and linearize the metric around
a Minkowski background
\begin{equation}
g_{\mu\nu} = \eta_{\mu\nu} + h_{\mu\nu}.
\end{equation}
Then coupling between the gravitational field and the photon becomes
\begin{equation}
{\cal L}_{\rm int} = -\frac{1}{4} \sqrt{-g} F_{\mu\nu} F^{\mu \nu} = 
\frac{1}{2} \gamma^{\mu\nu} F_{\mu\alpha} F_\nu^{\ \alpha} + {\cal O}(h^2)
\label{Lint}
\end{equation}
where 
\begin{equation}
\gamma_{\mu\nu} = h_{\mu\nu} - \frac{1}{4} \eta_{\mu\nu}h^\alpha_\alpha
\end{equation}
and the electromagnetic field strength is 
$F_{\mu\nu}=\partial_\mu A_\nu - \partial_\nu A_\mu$.  The coupling is quadratic in 
$A_\mu$ so that to lowest order in $h_{\mu \nu}$ photons are created in {\it pairs}.
The metric perturbation, $h_{\mu\nu}$, due to the cosmic string
energy-momentum tensor, $T_{\mu \nu}$, follows from the Einstein equations: in
Fourier space (denoted by tildes),
\begin{equation}
{\tilde \gamma}_{\mu\nu} = {\tilde h}_{\mu\nu} - \frac{1}{4} \eta_{\mu\nu} {\tilde h}^\alpha_\alpha \\
  = - \frac{16\pi G}{k^2} \left [ {\tilde T}_{\mu\nu}
          -\frac{1}{4} \eta_{\mu\nu} {\tilde T}_\alpha^{\alpha} 
                          \right ] \, .
                          \label{tildegamma}
\end{equation}

From the Nambu-Goto action, and using the conformal gauge 
\cite{VilenkinShellard}
\begin{equation}
T_{\mu\nu}(x) = \mu \int d^2\sigma
    ({\dot X}_\mu {\dot X}_\nu - X'_\mu X'_\nu) \delta^{4}(x-X)
    \label{Tstring}
\end{equation}
where $X^\mu (\sigma, t)$ is the string world-sheet.  A cosmic string loop trajectory can be written in terms of left- 
and right- movers
\begin{equation}
X^\mu (\sigma, t) = \frac{1}{2} [ a^\mu (\sigma_-) +
                         b^\mu (\sigma_+) ]
\label{loopX}
\end{equation}
where $\sigma_\pm = \sigma \pm t$,
and we will adopt world-sheet coordinates such that
\begin{eqnarray}
a^0 &=& -\sigma_- \ , \ \ b^0 = \sigma_+ \nonumber \\
|{\bm a}'| &=& 1 = |{\bm b}'|
\label{stringconstraints}
\end{eqnarray}
where primes denote derivatives with respect to the argument. Substituting (\ref{loopX}) into (\ref{Tstring}) yields
%
\begin{equation}
T_{\mu\nu}(x) = -\frac{\mu}{4} \int d\sigma_+ d\sigma_-
      ({a_\mu}' {b_\nu}' + {a_\nu}' {b_\mu}' ) \delta^4(x-X)
\end{equation}
which, when Fourier transformed, is
\begin{equation}
{\tilde T}_{\mu\nu}(k) = 
        -\frac{\mu}{4} \left ( I_{+,\mu} I_{-,\nu}
                       + I_{+,\nu} I_{-,\mu} \right ) \, .
                       \label{TstringFT}
\end{equation}
Here, for the periodic oscillations of a loop of length $L$
\begin{equation}
I_+^\mu = 
         \sum_{n=1}^\infty \delta \left( \frac{k^0L}{4\pi} -n \right )
          \int_0^L d\sigma_+ {b'}^\mu e^{-ik\cdot b/2}
\label{Iplus}
\end{equation}
\begin{equation}
I_-^\mu = 
         \sum_{n=1}^\infty \delta \left( \frac{k^0L}{4\pi} -n \right )
          \int_0^L d\sigma_- {a'}^\mu e^{-ik\cdot a/2} \, .
\label{Iminus}
\end{equation}
It will be important in the following to notice that as a result of the 
periodicity of the loop,
\begin{equation}
k\cdot I_\pm = 0 \, .
\label{useful1}
\end{equation}

\begin{figure}
  \includegraphics[width=0.25\textwidth,angle=-90]{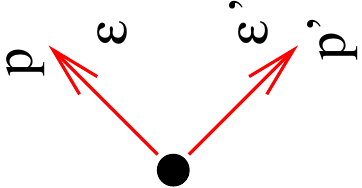}
  \caption{Feynman diagram showing two photon production from
           a classical string.}
  \label{fig:feynman}
\end{figure}

We can now calculate the amplitude for the pair creation of two outgoing photons
of momentum $p$ and $p'$, and polarisation $\epsilon$ and $\epsilon'$ respectively, where
\be
p^2 = 0 =p'^2 \, ; \qquad 
p\cdot \epsilon = 0 = p'\cdot \epsilon' \, .
\ee
This is given by the tree level process shown in 
Fig.~\ref{fig:feynman}, and on using equations (\ref{Lint}), (\ref{tildegamma}) and  (\ref{TstringFT}) we find
\begin{equation}
{\cal M}(p,p') = -\frac{4\pi G\mu}{k^2} I_+^\mu(k) I_-^\nu(k)  
                      Q_{\mu\nu}(p,\epsilon ; p', \epsilon')
\end{equation}
where momentum conservation imposes that
\begin{equation}
k=p+p',
\end{equation}
while
\ba
Q_{\mu\nu} &=& P_{\mu\nu}+P_{\nu\mu}
           - \frac{1}{2}\eta_{\mu\nu} P^\alpha_\alpha
\label{Qdef}
\\
P_{\mu\nu} &=& (p_\mu \epsilon_\alpha ^* - p_\alpha \epsilon_\mu^* )
             (p_\nu' {\epsilon '}^{\alpha *}  - 
                  {p'}^\alpha {\epsilon '}_\nu ^* )
 \label{Pdef}
\ea

The number of photon pairs produced in a phase space volume, the 
``pair production rate'', is given by ({\it e.g.} Sec.~4.5 of
\cite{PeskinSchroeder})
\begin{equation}
dN = \sum_{\epsilon , \epsilon '} \frac{d^3p}{(2\pi)^3} \frac{1}{2\omega}
     \frac{d^3p'}{(2\pi)^3} \frac{1}{2\omega'} |{\cal M}|^2
\label{eq:dN}
\end{equation}
where $\omega$ and $\omega'$ are the energies of the two outgoing photons,
while the energy emitted in the pairs is
\begin{equation}
dE = \sum_{\epsilon , \epsilon '} \frac{d^3p}{(2\pi)^3} \frac{1}{2\omega}
 \frac{d^3p'}{(2\pi)^3} \frac{1}{2\omega'} (\omega +\omega') |{\cal M}|^2 \, .
\label{eq:dE}
\end{equation}
Clearly the crucial relevant quantity is
\ba
|{\cal M}|_{\rm tot}^2 &\equiv& \sum_{\epsilon,\epsilon'}  |{\cal M}|^2 
 \nonumber \\
&& \hskip -0.5 in 
= \left ( \frac{4\pi G\mu}{k^2} \right )^2 \sum_{\epsilon,\epsilon'}  
{I_+^\alpha}I_+^{\mu *} {I_-^\beta} I_-^{\nu *}
Q_{\alpha\beta} Q^*_{\mu\nu} \, .
\label{Mtot}
\ea
Note that in this paper we
only study the total radiation rate from a cosmic string loops:
a discussion of any polarization signatures is left for subsequent
work.

After substitution of $Q_{\mu \nu}$ given in Eq.~(\ref{Qdef}), the sum over photon polarizations in (\ref{Mtot}) can be simplified through the replacement
\be
\sum_{\epsilon,\epsilon'} \epsilon^*_\mu \epsilon_\nu 
\rightarrow -\eta_{\mu \nu} \ ,
\label{util}
\ee
provided certain conditions hold \cite{PeskinSchroeder}. More specifically, let us define ${\cal M}_{\rho \sigma \mu\nu}$ via
\be
Q_{\mu\nu} = \epsilon^{\rho *}{\epsilon '}^{\sigma *}
                     {\cal M}_{\rho \sigma \mu\nu}
\ee
so that
\ba
{\cal M}_{\rho\sigma\mu\nu} &=& N_{\rho\sigma\mu\nu}+N_{\rho\sigma\nu\mu}
          -\frac{1}{2} \eta_{\mu\nu} N_{\rho\sigma~ \alpha}^{~~\alpha}
\label{M1}
\\
N_{\rho\sigma\mu\nu}&\equiv& (p_\mu \eta_{\alpha\rho}-p_\alpha \eta_{\mu\rho})
      (p_\nu' \delta^\alpha_\sigma   - 
                  {p'}^\alpha \eta_{\sigma \nu}  )
                  \label{N1}
\ea
Then the required condition \cite{PeskinSchroeder} is that
\be
p^\rho {\cal M}_{\rho\sigma\mu\nu}= 0 =
                              {p '}^\sigma {\cal M}_{\rho\sigma\mu\nu}.
\label{pdotM}
\ee
However, it is straightforward to check that this condition is satisfied, since it is an immediate consequence of the definition of $ {\cal M}_{\rho\sigma\mu\nu}$ in (\ref{M1})-(\ref{N1}). 

Then, after quite a bit of algebra and on using (\ref{util}), we find 
that $|{\cal M}|_{\rm tot}^2$ is given by
\ba
|{\cal M}|_{\rm tot}^2 &=& 
\left ( \frac{2\pi G\mu}{p\cdot p'} \right )^2 
\biggl [ ~
8 |p\cdot I_+ |^2 |p\cdot I_-|^2 
              \nonumber \\
&& 
\hskip -0.75in
+4 p\cdot p' 
\left\{ ~ |p\cdot I_+|^2 |I_-|^2 + |p\cdot I_-|^2 |I_+|^2  \right.
\nonumber \\
&& 
\hskip -0.3in
  +  p\cdot I^*_+ ~p\cdot I_- ~I_+\cdot I^*_- 
   +p\cdot I_+ ~ p\cdot I^*_- ~ I^*_+\cdot I_- 
\nonumber \\
&& 
\hskip -0.3in
     - p\cdot I_+ ~p\cdot I_- ~I^*_+\cdot I^*_-
     \left. - p\cdot I^*_+ ~p\cdot I^*_- ~I_+\cdot I_- ~
\right \}
\nonumber \\
&& \hskip -0.7 in
+ 2 (p\cdot p')^2 \left \{ |I_+|^2|I_-|^2 + |I^*_+\cdot I_-|^2
        - |I_+\cdot I_-|^2 \right \} ~ \biggr ]
\label{Mtot2}
\ea
where
$|I_\pm|^2 \equiv I^{*\mu}_\pm I_{\pm\mu}$, and in order to simplify the result we have made extensive use of $p\cdot I_\pm = - p'\cdot I_\pm$ which follows (\ref{useful1}) since $k=p+p'$.  Finally, we have expressed the answer
 in powers of $k^2 = 2p\cdot p'$: this will be important later when we will see that the $k^2 \rightarrow 0$ limit plays a crucial role.

\section{Evaluation of $|{\cal M}|_{\rm tot}^2$}
\label{sec:integrals}

In order to calculate the energy radiation in photon pairs, we need to evaluate $|{\cal M}|_{\rm tot}^2$, where the dynamics of cosmic string loops enters Eq.~(\ref{Mtot2}) through the integrals $I_\pm(k)$. These
integrals, defined in (\ref{Iplus})-(\ref{Iminus}), also occur in the calculation of other forms of radiation
from strings and have been discussed in the past 
({\it e.g.}~\cite{Vilenkin:1986zz,Damour:2001bk}).  
There is, however, a key difference between Aharonov-Bohm (AB) radiation and other
forms of radiation from strings that are commonly studied: namely
AB radiation involves {\it two} particle final states. 
As a result, the kinematics of the problem is potentially
more complicated. 

Generally, however, it is well known that $I_\pm(k)$ decay exponentially 
with $k^0 L$, where $L$ is the length of the loop, unless either the 
phase in these integrals has a saddle point on the real line, or there 
is a discontinuity in the integrand due to kinks in $b'^\mu$ and/or 
$a'^\mu$ (see {\it e.g.~} Ch.~6 of Ref.~\cite{BenderOrszag}). 
Below we study these cases in turn, and we will see that despite the two-particle nature of the final state, a saddle point in both $I_+^\mu$ and $I_-^\mu$ corresponds to a 
cusp on the string -- namely $|\dot{\vec X}|=1$; a saddle point in 
one of the integrals and a discontinuity in the other occurs at a kink.  
Finally, when two kinks collide, there is a discontinuity in both 
integrands. In all three possibilities -- cusp, kink, kink-kink
collision -- $I_{\pm}^\mu$ decay as a powerlaw, 
$(k^0 L)^{-q}$, where the index $q$ will be determined below.

\subsection{Saddle points and cusps}

As a first step in evaluating $I_\pm$, 
we establish certain relations between the momenta of particles 
emitted when there are saddle points in these integrals.
On recalling that both $p^2=0=p'^2$, let us write
\ba
p &=& (\omega, {\vec p}) \; = \; \omega(1, \hat{\vec p}) \; \equiv \; \omega \hat{p}
\nn
\\
p' &=& (\omega', {\vec p}')\; = \; \omega'(1, \hat{\vec p}') \; \equiv \; \omega' \hat{p}'
\label{pp'}
\ea
where $|\hat{\vec p}|=1=| \hat{\vec p'}|$.
Since $k=p+p'$ then
\begin{equation}
k^2 = 2 p\cdot p' = 2\omega\omega' (1-{\hat {\vec p}}
                                 \cdot {\hat {\vec p}}')\, .
\label{k22pp'}
\end{equation}
We also define 
\begin{equation}
k = (\Omega, {\vec k}) = (\omega + \omega', {\vec p}+{\vec p}').
\end{equation}

Next consider the integral $I_+ (k)$ in Eq.~(\ref{Iplus}) when
there is a saddle point. This implies that there is a point
such that
\begin{equation}
k \cdot b' = 0 
\label{sad}
\end{equation}
where from the gauge conditions Eq.~(\ref{stringconstraints}) $b'^2=0$. Thus at a saddle point
\begin{equation}
{\hat k} = {b'},
\; \;  {\rm where}\; \; \;  {\hat k} \equiv \frac{k}{\Omega}
\label{kisb}
\end{equation}
so that
\begin{equation}
k^2 = \Omega^2 {\hat k}^2 = \Omega^2 {b'}^2 =0.
\label{k2is0}
\end{equation}
Thus $k^2$ is {\it null}.
Furthermore, from Eq.~(\ref{k22pp'}),
\begin{equation}
{\hat p}  = {\hat p}^{\prime} = {\hat k} = b' \ ,
\label{pisk}
\end{equation}
where $\hat{p}$ and $\hat{p'}$ were defined in Eq.~(\ref{pp'}).
%

Apart from a sign, all the above goes through for a saddle point in $I_-$.
Again $k^\mu$ must be light-like, but now 
$k\cdot a'=0 = \Omega (-1+{\hat \veck}\cdot \veca')$ 
(since $a^0=-\sigma_-$).  Thus
\begin{equation}
\hat{p} = \hat{p}^{\prime}=\hat{k} = -a'.
\label{piskb}
\end{equation}

Evaluation of the integrals around the saddle points, in the 
$\Omega L \gg 1$ limit can then be carried out in the standard way 
(see {\it e.g.~}\cite{BenderOrszag}) and, using the kinematic relations 
given above, leads to 
\begin{eqnarray}
I_{+,n}^{\rm saddle} &=& A_+ L \frac{b_s^{\prime}}{(\Omega L)^{1/3}} + 
     i B_+ L^2 \frac{b_s^{\prime \prime}}{(\Omega L)^{2/3}} + \ldots
\label{ipsaddle}
\\
I_{-,n}^{\rm saddle}  &=& A_- L \frac{a_s^{\prime}}{(\Omega L)^{1/3}} + 
     i B_- L^2 \frac{a_s^{\prime \prime}}{(\Omega L)^{2/3}} +\ldots \, .
\label{imsaddle}
\end{eqnarray}
where the subscript $n$ on $I_\pm$ refers to the $n^{\rm th}$ term
in the sum in Eqs.~(\ref{Iplus}) and (\ref{Iminus}), and the
subscript $s$ on $a^\mu$ and $b^\mu$ refers to evaluation at a saddle 
point. The delta functions in Eqs.~(\ref{Iplus}) and (\ref{Iminus})
enforce $\Omega L = 4\pi n$. We have dropped an overall phase factor 
which is irrelevant because it is the square of the amplitude that 
gives a rate.
The coefficients can be evaluated explicitly;
\begin{eqnarray}
A_+&=&  \left( \frac{12 }{L^2 |b''|^2} \right)^{1/3} 
         \frac{2 \pi}{ 3 \Gamma(2/3)},
\nonumber \\
B_+ &=& \left( \frac{12 }{L^2 |b''|^2} \right)^{2/3} 
               \frac{1}{\sqrt{3}}{ \Gamma(2/3)}
\end{eqnarray}
and $A_-$ and $B_-$ are given by identical expressions except that
$b''$ is replaced by $a''$. 

A cusp on a string loop occurs when $|\dot{\vec X}|=1$ and hence, from (\ref{loopX}), when ${\vec a}' = - {\vec b}'$. However, from Eqs.~(\ref{pisk}) and (\ref{piskb}) this condition requires a saddle point
contribution to both $I_+$ and $I_-$. In the vicinity of the 
beam of the cusp, from Eqs.~(\ref{kisb}), (\ref{pisk}), we get
${\hat {\vecp}} \sim {\vec b}_c' = -{\vec a}_c'$, and similarly 
for $p'$. Therefore from (\ref{ipsaddle}) and (\ref{imsaddle}), 
we estimate
\begin{equation}
p\cdot I_\pm^{\rm saddle} \sim {\cal O}(k^2) \ ,  \ \ 
p'\cdot I_\pm^{\rm saddle} \sim {\cal O}(k^2) .
\label{scal}
\end{equation}

This is an important result: consider $|{\cal M}|_{\rm tot}^2$ given in  Eq.~(\ref{Mtot2}). In the beam of the cusp $k^2 =2p\cdot p' \to 0$ (Eq.~(\ref{k2is0})),  so one might worry that  Eq.~(\ref{Mtot2}) diverges due to the overall factor of $1/(k^2)^2$.
However, this divergence is rendered harmless by the scaling in Eq.~(\ref{scal}). Indeed, in the $k^2=0$ limit,  only the last line in 
Eq.~(\ref{Mtot2}) gives the dominant contribution to the emission 
from the cusp:
\ba
(|{\cal M}|_{\rm tot}^2)_{\rm cusp} &\to& 
 ( 2\pi G\mu )^2 \nonumber \\
 &&\hskip -0.8 in
\times 2 [\, |I_+|^2|I_-|^2 + |I^*_+\cdot I_-|^2 - |I_+\cdot I_-|^2 \, ] 
\label{Mcusp2}
\ea
with $I_\pm$ in Eqs.~(\ref{ipsaddle}) and (\ref{imsaddle}) (we have dropped the label ``saddle'').
The other terms in Eq.~(\ref{Mtot2}) all contain factors such
as $|p\cdot I_\pm|$ and are higher order in $k^2$.  

The above analysis assumes that $\veck$ is in the direction of the
cusp. If $\veck$ is at some small angle, $\theta_+$, to $\vecb '$, 
we can write ${\hat \veck} \cdot \vecb' = \theta_+^2/2$ and repeat
the above analysis as in \cite{Vilenkin:1986zz}. The estimate is valid 
for
\be
\theta_+ \le \theta_{m,+} \equiv
\left ( \frac{4L|\vecb '' |^2}{\sqrt{3}\Omega} \right )^{1/3} 
\label{theta+}
\ee
Similarly in the case of $I_-$
\be
\theta_- \le \theta_{m,-} \equiv
\left ( \frac{4L|\veca '' |^2}{\sqrt{3}\Omega} \right )^{1/3}
\label{theta-}
\ee
This estimate assumes $k^2=0$ but it holds even if $k$ is perturbed 
so that it is not precisely null. 
To summarize, the estimate (\ref{Mcusp2}) holds in a cone of opening 
angle $\theta_+\sim\theta_-\sim (\Omega L)^{-1/3}$.

\subsection{Discontinuities and kinks}

Next we find the contribution of a discontinuity to the
integrals $I_\pm$. On expanding the integrands on both sides of the
discontinuity, which is say in $a'$ at $\sigma_-=u_k$, one can extract the dominant contribution;
\ba
I_-^{\rm disc} &\sim& \int^{u_k} d\sigma_- ~ a'_- 
           e^{-ik\cdot (a_k + a'_- (\sigma_- -u_k) + \ldots)}
        \nonumber \\
 &&  \hskip 0.2 cm
+ \int_{u_k} d\sigma_- ~ a'_+ 
          e^{-ik\cdot (a_k + a'_+ (\sigma_- -u_k)+\ldots )}
       \nonumber \\ 
&=& -\frac{2}{i\Omega} 
    \left ( \frac{a'_+}{{\hat k}\cdot a'_+}
           -\frac{a'_-}{{\hat k}\cdot a'_-} 
    \right ) e^{-ik\cdot a_k}
\label{I-disc}
\ea
where $a'_\pm$ refers to the value of $a'$ on either side of the discontinuity,
and $a_k = a(\sigma_-=u_k)$. 
Similarly we can find $I_+^\mu$, and the result is
\be
I_+^{\rm disc} \sim -\frac{2}{i\Omega} 
    \left ( \frac{b'_+}{{\hat k}\cdot b'_+}
           -\frac{b'_-}{{\hat k}\cdot b'_-} 
    \right ) e^{-ik\cdot b_k} \, .
\label{I+disc}
\ee
It is important to observe that $I_\pm^{\rm disc} \sim \Omega^{-1}$
decays faster with frequency than  $I_\pm^{\rm saddle}$
(Eqs.~(\ref{ipsaddle}), (\ref{imsaddle})).

The estimates in eqns.~(\ref{I-disc}) and (\ref{I+disc}) preserve 
the relation $k \cdot I_\pm =0$. Hence {\it if} $k^2 \to 0$ then
${\hat p}^\mu = {\hat p}^{\prime\mu} = {\hat k}^\mu$ and again 
$p\cdot I_\pm^{\rm disc} \to 0$ and $p'\cdot I_\pm^{\rm disc} \to 0$. 
These relations are important to see that the expression for the invariant 
matrix element in Eq.~(\ref{Mtot2}) is not singular in the $k^2 \to 0$ 
limit. The case when ${\hat k}\cdot b'_\pm =0$ or ${\hat k}\cdot a'_\pm =0$ 
is very special because now there is a saddle point in addition to a discontinuity
and we shall not consider its consequences.

While a saddle point in both $I_+^\mu$ and $I_-^\mu$ corresponds to a 
cusp on the string, a saddle point in 
one of the integrals and a discontinuity in the other occurs at a kink.  
Thus the dominant contribution to photon production 
from a kink takes place exactly when $k^2 \to 0$, namely in the forward direction, 
when ${\hat p}$ and ${\hat p}'$ are collinear. 
In this limit, 
\be
|p\cdot I^{{\rm disc}}_\pm|^2 \sim O(k^2), \qquad
|p'\cdot I^{{\rm disc}}_\pm|^2 \sim O(k^2)
\label{termorderkinks}
\ee
which should be compared to Eq.~(\ref{scal}) for a cusp.
This can be most clearly seen by writing,
for example,
\begin{eqnarray}
p\cdot I_-^{\rm disc} &\sim&
p\cdot \left ( \frac{a'_+}{k\cdot a'_+}
              -\frac{a'_-}{k\cdot a'_-} 
       \right ) 
\nonumber \\
&& 
\hskip -0.7 in
= \left \{ \frac{\omega\omega'}{\Omega^2}
\frac{[({\hat \vecp} \cdot {\vec a}'_+) {\vec a}'_- - 
       ({\hat \vecp} \cdot {\vec a}'_-) {\vec a}'_+ ]}
     {{\hat k}\cdot a'_+ ~ {\hat k}\cdot a'_-}
\right \}
  \cdot ( {\hat \vecp}' - {\hat \vecp} )
\end{eqnarray}
Therefore $p\cdot I_-^{\rm disc} = O( | {\hat \vecp}' - {\hat \vecp} | )$
from which (\ref{termorderkinks}) follows since
$\omega\omega' ( {\hat \vecp}' - {\hat \vecp} )^2 = k^2$
(see Eq.~(\ref{k22pp'})).

\subsection{No saddle point or discontinuities}

If a loop has neither cusps or kinks, then $I_\pm$ decay exponentially with 
$\Omega L$. In that case the energy radiated by the loop per unit unit time, 
which is proportional to $|{\cal{M}}|^2_{\rm tot}$, also decays exponentially 
as  $\dot{E}_n \propto e^{- \alpha n}$ where 
$n = \Omega L/(4\pi)$ is the harmonic number and $\alpha$ is a 
coefficient. In other words, a loop with no kinks or cusps will 
radiate a {\it finite} amount of energy. We have checked explicitly 
that this is the case by considering the radiation from a chiral 
cosmic string loop for which there are no cusps and kinks.
On the other hand, as we now show, the radiation from a cusp or a 
kink diverges and needs to be cut off due to the thickness
of the string.

\section{Power Emitted}
\label{power}

We now evaluate the power emitted from cusps, kinks and kink-kink collisions.
We divide (\ref{eq:dE}) by $L$ to get the average power radiated in the 
$n^{\rm th}$ harmonic 
\begin{eqnarray}
{\dot E}_n &=& \frac{2\pi}{L^2}  n
 \int \frac{d^3p}{(2\pi)^3} \frac{1}{2\omega}
 \int \frac{d^3p'}{(2\pi)^3} \frac{1}{2\omega'}  \nonumber \\
 && \hskip 0.2 cm \times 
  |{\cal M}|_{n,{\rm tot}}^2
           \delta(n-(\omega + \omega ')L/4\pi  ) \ ,
\label{powern}
\end{eqnarray}
while the total power is the sum over all harmonics. 
Note that $\Omega = \omega+\omega' = 4\pi n/L$.


\subsection{Emission from cusp}
\label{cuspemission}

In order to get the power radiated from a cusp, we first insert 
the expressions for $I_\pm^{\rm saddle}$ given in Eqs.~(\ref{ipsaddle}) and 
(\ref{imsaddle}) into Eq.~(\ref{Mcusp2}), remembering to include the
sum and delta functions in Eqs.~(\ref{Iplus}) and (\ref{Iminus}).
Not all the terms in $I_\pm^{\rm saddle}$ contribute since the string constraint equations (\ref{stringconstraints})
imply
\be
|a'|^2 = 0 = |b'|^2 \ , \ \ 
a'\cdot a'' = 0 =b'\cdot b''
\ee
and, in addition, since $\veca'=-\vecb'$ at the cusp,
\be
b_c'\cdot a_c'' = 0 = a_c'\cdot b_c'' 
\ee
(where the subscript $c$ denotes cusp).
Then to leading order we find 
$|I_+^*\cdot I_-| = |I_+\cdot I_-|$ and 
the non-vanishing contributions in (\ref{Mcusp2}) 
come from the $|I_+|^2|I_-|^2$ term which is proportional
to $|a_c''|^2 |b_c''|^2$, so that
\ba
(|{\cal M}|_{n,\rm tot}^2)_{\rm cusp}& =& 
2 ( 2\pi G\mu )^2 |I_+|^2|I_-|^2 
\nn
\\ &&
\hskip -0.5 in
 \sim (G\mu)^2 L^4\frac{1}{ n^{8/3}}  
|{\veca}''_c L|^{-2/3} |{\vecb}''_c L|^{-2/3}
\label{ccuu}
\ea
where we ignored an overall numerical factor of order 1.

In order to calculate (\ref{powern}) and estimate ${\dot E}_n$ we next rescale the momenta $\vecp$ and $\vecp'$ by $4\pi n/L$. Then
\be
 \int \frac{d^3p}{(2\pi)^3} \frac{1}{2\omega} =
\frac{1}{2(2\pi)^3} \left ( \frac{4\pi n}{L} \right )^2 
          \int d{\bar \omega} {\bar \omega} \int d^2 {\hat \vecp}
\label{phasevolume}
\ee
where ${\bar \omega} = L |{\vecp}|/4\pi n$. The integration
over ${\bar \omega}$ gives an order 1 numerical factor. The 
integration over the direction ${\hat \vecp}$ is estimated as
\ba
\int d^2 {\hat \vecp} d^2 {\hat \vecp}' &\sim&
 \pi \theta_{m,+}^2  ~ \pi \theta_{m,-}^2 \nonumber \\
&\sim&   |{\veca}''_c L|^{4/3} |{\vecb}''_c L|^{4/3} n^{-4/3}
\label{phatintegration}
\ea
where we have used $\theta_{m,\pm}$ given in Eqs.~(\ref{theta+})
and (\ref{theta-}). 

Now putting together results in
Eqs.~(\ref{powern}), (\ref{ccuu}), and (\ref{phatintegration}) 
we finally obtain
\be
{\dot E}_n \bigg |_{\rm cusp} \approx 
 \left ( \frac{G\mu}{L} \right )^2
  |{\veca}''_c L|^{2/3} |{\vecb}''_c L|^{2/3}.
\label{edotncusp}
\ee
The most important feature of this 
estimate \footnote{We have dropped factors of $2$ and $\pi$ in this 
estimate; we have checked that including them gives a factor ${\cal O}(1)$.}
 is that it does not depend on the harmonic number $n$: all factors of $n$ 
have cancelled!  This result was noted earlier in \cite{JonesSmith:2009ti} 
in the context of scalar radiation and in \cite{Chu:2010zzb} 
in the context of fermion radiation, though for specific loop trajectories. 
Our analysis here is more general. Additional features 
of interest are the $1/L^2$ dependence that will result
in larger radiation from smaller loops, and the dependence 
on the (dimensionless) cusp acceleration vectors, $L \veca''_c$ 
and $L \vecb''_c$. 

The $n$-independent spectrum of gravitational Aharonov-Bohm 
radiation implies that photons of arbitrarily high frequencies
will be emitted from cusps. Naively this would imply a divergence
in the emitted power. However, the Nambu-Goto action ignores
the thickness of a field theory string, and this suggests
a cutoff for the highest harmonic that can be emitted --
the highest harmonic emitted from a cusp is the string width
in the rest frame of the string. To work out the Lorentz
factor we start out by noting that the saddle point analysis
gives contributions from the region around the cusp of size
$|k\cdot a'| < 1$ (in the case of $I_-$) where the phase
in the integrand is not oscillating rapidly. 
(We follow the discussion in \cite{Vachaspati:2009kq}.)
This gives a time and length interval on the string world-sheet 
\be
|\Delta t|, |\Delta \sigma| 
< \frac{L}{(\Omega L)^{1/3}}\ .
\label{beamduration}
\ee
In this region
\be
1-{\dot{\vec x}}^2 
            \sim \frac{(\Delta\sigma)^2}{L^2} 
 \sim (\Omega L)^{-2/3}
\ee
Requiring that
$\Omega$ be less than the Lorentz boost
factor times the inverse string width, and ignoring numerical
factors we get
\be
\Omega < M \sqrt{ML}
\ee
where $M$ is the mass scale associated with the string.
In terms of the harmonic number, this gives a cutoff
\be
n_c \sim (ML)^{3/2} \ .
\ee
Hence Eq.~(\ref{edotncusp}) applies for $n \le n_c$.

The result in Eq.~(\ref{edotncusp}) gives the power emitted in 
the $n^{\rm th}$ harmonic. Observationally, it is the energy 
flux at the observer's location that is relevant. To get
this flux we divide ${\dot E}_n$ by the cross-sectional area of
the beam emitted from the cusp at a distance $r$ from the cusp
\be
\frac{{\dot E}_n}{\pi \theta_m^2 r^2} \sim
\frac{1}{r^2} \left ( \frac{G\mu}{L} \right )^2 n^{2/3} 
\ee
where we have estimated the dimensionless cusp acceleration
to be order unity, and the width of the beam as in 
(\ref{theta-}).

The flux grows with $n$ suggesting that it is advantageous to
observe at high frequencies. However, the beam at higher $n$ is 
narrower and hence event rates are lower at high frequencies. 
These are observational issues that we will return to in future
work.


\subsection{Emission from kink}
\label{kinkemission}

Now we estimate the power emitted from a kink (Eq.~(\ref{powern})),
say when there is a discontinuity in ${\vec a}'$ that contributes
to $I_-$, and a saddle point that contributes to $I_+$. Then 
$I_- = I_-^{\rm disc}$ is given by (\ref{I-disc}) and 
$I_+ = I_+^{\rm saddle}$ by (\ref{ipsaddle}). 
Note that, for a given $k^\mu$, there will be a saddle point 
contribution to $I_+$ if there exists a $\sigma_+$ such that 
$k\cdot b'(\sigma_+) =0$. Since ${\vec b}'(\sigma_+)$ describes
a curve parametrized by $\sigma_+$, there is a saddle point 
contribution for every ${\vec k}$ in the direction of a point
on the ${\vec b}'$ curve. Hence the saddle point contribution 
applies to emission along a curve of directions.

Since there is a saddle point in $I_+$, the kinematic relations discussed 
in Sec.~\ref{cuspemission} (Eqs.~(\ref{pp'})-(\ref{pisk})) apply.
Noting the estimate in Eq.~(\ref{termorderkinks}), we find that
the dominant terms in (\ref{Mtot2}) in the $\Omega L \gg 1$ limit 
are given by
\be
(|{\cal M}|_{n, {\rm tot}}^2)_{\rm kink} \to 
2 \frac{(2\pi G\mu)^2}{p\cdot p'} |I_+|^2
\left [ 2 |p\cdot I_-|^2 +  p\cdot p' |I_-|^2 \right ]
\label{M2kfinal}
\ee
where we have also used $|I_+^*\cdot I_-| = |I_+\cdot I_-|$. 
Now we can substitute $I_-$ from Eq.~(\ref{I-disc}) and
$I_+$ from (\ref{ipsaddle}). Ignoring numerical factors,
the result can be written as 
\be
(|{\cal M}|_{n,{\rm tot}}^2)_{\rm kink} \sim (G\mu)^2 L^4  \frac{1}{n^{10/3}}
    ({\rm shape\ factors})
    \label{kki}
\ee 
where the ``shape factors'' include various scalar products
that depend on $a'_\pm$ -- the shape of the loop -- and the 
directions of the momenta. 
The shape factors can be written down using the expression 
in (\ref{M2kfinal}) but the result is not illuminating.  
Expression (\ref{kki}) should be compared with the analogous 
one, Eq.~(\ref{ccuu}), for a cusp.

We now turn to the phase space integrals in (\ref{powern}). 
The phase space volume is given in (\ref{phasevolume}) except 
that, unlike
in the case of the cusp, there is a whole curve of saddle points 
that are relevant, and the emission is along a curve of directions.
Hence the integration over momentum direction is estimated as
\be
\int d^2 {\hat \vecp} \sim 2 \pi \theta_{m,\pm}
\label{phatintegrationkink}
\ee
where $\theta_{m,\pm}$ are given in (\ref{theta+}), (\ref{theta-}).

Putting together all the pieces in Eq.~(\ref{powern}) and 
using (\ref{phasevolume}), we obtain
\be
{\dot E}_n\bigg |_{\rm kink} \sim \left ( \frac{G\mu}{L} \right )^2 
    ({\rm shape\ factors}) 
\label{Enkink}
\ee
As in the cusp case, an important feature of this result is 
that it is independent of the frequency of emission. 

We now denote the cutoff in harmonic number by $n_k$, and estimate it
by the thickness of the string. What is different from the cusp is 
that the velocity of the string at the kink is not ultra-relativistic
and there is no corresponding large Lorentz boost factor. Therefore
the estimate in (\ref{Enkink}) applies for
\be
n < n_k \sim ML \ .
\ee
%
%
The energy flux from a kink at distance $r$ is now given by
\be
\frac{{\dot E}_n}{\theta_m r^2} \sim
\frac{1}{r^2} \left ( \frac{G\mu}{L} \right )^2 n^{1/3} \ .
\ee

\subsection{Emission from kink-kink collision}
\label{kinkkinkcollisionemission}

This is the situation in which both $I_-$ and $I_+$ get contributions 
from discontinuities in $a'$ and $b'$ respectively. Therefore,
in the general formula (\ref{Mtot2}) we have to insert 
(\ref{I-disc}), (\ref{I+disc}). This gives
\ba
|{\cal M}|_{\rm tot}^2 &\to& 
2 \left ( \frac{2\pi G\mu}{p\cdot p'} \right )^2 
\biggl [
4 |p\cdot I_+ |^2 |p\cdot I_-|^2 
              \nonumber \\
&& \hskip -0.3in
+2 (p\cdot p')^1 
\{ |p\cdot I_+|^2 |I_-|^2 + |p\cdot I_-|^2 |I_+|^2 \}
\nonumber \\
&& \hskip 0.6 in
+  (p\cdot p')^2 |I_+|^2|I_-|^2 \biggr ] \ .
\label{Mtot2kk}
\ea
The estimate in Eq.~(\ref{termorderkinks}) shows that all the terms
are non-singular if we take $k^2 \to 0$. 

To evaluate the power radiated from a kink-kink collision, we
see from Eqs.~(\ref{I+disc}) and (\ref{I-disc}) that $I_\pm$ are
$O(1/n)$. Hence $|{\cal M}|^2 \sim n^{-4}$. Also, since there is
no beaming in the kink-kink collision
\be
\int d^2 {\hat \vecp} \sim \pi \ .
\label{phatintegrationkk}
\ee
Then, putting together factors in Eq.~(\ref{powern}) and 
using (\ref{phasevolume}), we obtain
\be
{\dot E}_n\bigg |_{\rm k-k} \sim \left ( \frac{G\mu}{L} \right )^2 
    ({\rm shape\ factors}) 
\label{Enkk}
\ee
exactly as in the estimate for the cusp and the kink, though
the shape factors are different in all three cases and in this 
kink-kink case they may vary with direction of the momenta. 
Once again, the result is independent of the harmonic number 
$n$ and holds up to $n_k = ML$, as in the kink case.

The energy flux from a kink-kink collision at distance $r$ is
\be
\frac{{\dot E}_n}{r^2} \sim
\frac{1}{r^2} \left ( \frac{G\mu}{L} \right )^2 \ .
\ee

The estimate (\ref{Enkk}) is due to contribution of the 
discontinuities at fixed values of both  $\sigma_+$ and $\sigma_-$.
Thus the emission is coming from a single point on the string
at one instant of time. This corresponds to the point where
a left-moving kink and a right-moving kink collide. Hence the
temporal duration of the burst is set by the string thickness.
On the other hand, emission at a frequency $\omega$ cannot be
temporally resolved in a time interval less than $\sim \omega^{-1}$.
For this reason, the observed burst duration at frequency $\omega$ 
is set by $\omega^{-1}$.

The total energy emitted in harmonic $n$ during a kink-kink 
collision can be estimated from (\ref{Enkk}), which is the
emitted power averaged over a time period $L$. Hence the
total energy emitted in the $n^{\rm th}$ harmonic in one
kink-kink collision is 
\begin{equation}
{E}_n\bigg |_{\rm k-k} \sim \frac{(G\mu)^2}{L}   
    ({\rm shape\ factors}) .
\end{equation}
The continuum version may now be written as
\begin{equation}
\frac{dE}{d\omega} \bigg |_{\rm k-k} \sim (G\mu)^2 \psi_a \psi_b  
    ({\rm other\ shape\ factors}) .
\label{onekk}
\end{equation}
where the ``sharpness'' $\psi_a$ is defined by \cite{Copeland:2009dk}
\begin{equation}
\psi_a = \frac{1}{2} (1 - {\veca}_+'\cdot {\veca}_-')
 = \frac{1}{4}({\veca}_+' - {\veca}_-')^2
\end{equation}
and similarly for $\psi_b$. In Eq.~(\ref{onekk}) we have pulled out
factors of the sharpness since the result must vanish if the sharpness
vanishes. The ``other shape factors'' will in general also depend on the 
direction of emission.

\section{Conclusions}
\label{conclusions}

In this paper we have calculated the flux of photons from cosmic strings. In general
this falls off exponentially with harmonic number $n$. However, as we have shown, the
power emitted from cusps, kinks and kink-kink collisions does not fall 
off with $n$ -- rather, it is $n$-{\it independent}.  Thus the emission from
these features on the string dominates over the emission from the rest of
the string, at least at high frequencies. If we denote the differential energy flux 
at frequency $\omega_0$ by $F$ {\it i.e.}
\be
F \equiv \frac{d^3E}{{\rm d} t \; {\rm d}\omega_0 \; {\rm d}\Omega_{\rm s}} 
\ee
where $\Omega_{\rm s}$ denotes solid-angle,
then our results can be summarized as follows:
\ba
&&
\hskip -0.8 cm
F_{\rm cusp} \approx 
\frac{(G\mu)^2}{L} (\omega_0 L)^{2/3}, \ \ 
\Omega_{\rm s} < (\omega_0 L)^{-2/3}
\\
&&
\hskip -0.5 cm
F_{\rm kink} \approx 
\frac{(G\mu)^2}{L} (\omega_0 L)^{1/3}, \ \ 
\theta < (\omega_0 L)^{-1/3}
\\
&&
F_{\rm k-k} \approx \frac{(G\mu)^2}{L} \ .
\ea
The cusp emits a beam within a solid angle, the kink emits 
along a curve, while the kink-kink emission is in all directions. 
The duration of the cusp and kink beams is given by 
Eq.~(\ref{beamduration}), while the duration of the kink-kink 
radiation is given by the wavelength at which the emission is
observed. Also, the cusp radiates at frequencies $\omega_0 < M\sqrt{ML}$
whereas the kink and kink-kink collisions radiate for $\omega_0 < M$,
where $M$ is the string scale.

Our results so far provide the emission characteristics from
certain features on strings. Now we briefly discuss the cumulative
effect of having many such features on a given loop of string. The cusp
and kink emissions are beamed and this makes the analysis more
involved. However, the emission due to kink-kink collisions is
not beamed and is easier to estimate.

Eq.~(\ref{onekk}) gives the energy emitted from a single kink-kink
collision. To get the energy emitted from a string segment, we need 
to sum over all the kink-kink collisions occurring on that string
segment of length $\Delta l$ in an interval of time $\Delta t$
\begin{equation}
\frac{dE}{d\omega} \sim (G\mu)^2 
     \int d\psi_a ~ d\psi_b ~ \psi_a \psi_b 
                      \frac{dn_a}{d\psi_a} \frac{dn_b}{d\psi_b}
                      ~ \Delta l ~ \Delta t
\label{Eall}
\end{equation}
where $n_a (\psi_a, t)$ is the number of kinks of sharpness $\psi_a$
at time $t$ per unit length of string, and similarly for $n_b$.

We will consider emission from a loop of string that formed at time
$t_f$ by breaking off a long string. The loop inherits a large
number of (shallow) kinks from the long string and
from Ref.~\cite{Copeland:2009dk} we can write
\begin{equation}
\int d\psi_a \psi_a \frac{dn_a}{d\psi_a} \sim 
    \frac{1}{t_f} \left ( \frac{t_f}{t_*} \right )^\alpha \ .
\label{avgpsi}
\end{equation}
The exponent $\alpha$ is $\sim 0.7$ in the radiation-dominated epoch.
In the matter-dominated epoch, strings contain all the kinks accumulated
until the epoch of matter-radiation equality, $t_{\rm eq}$, and from
then on the scaling in (\ref{avgpsi}) has $\alpha \sim 0.4$ \cite{CopelandKibble}.
The time $t_*$ denotes the epoch at which frictional effects on strings 
became unimportant. 
Hence Eq.~(\ref{Eall}) can be written as
\begin{equation}
\frac{dP_\omega}{dl d\omega} \sim \frac{(G\mu)^2}{t_f^2 } 
             \left ( \frac{t_f}{t_*} \right )^{2 \alpha} 
\label{Pall}
\end{equation}
where $P_\omega$ denotes the power emitted at frequency $\omega$.

We now obtain some numerical estimates, leaving a detailed analysis
for future work. The photons emitted from loops deep into the radiation
epoch will get thermalized. Hence the emission from loops in the
post-recombination era is most relevant for direct observation. 
Loops at the epoch of recombination could have been produced in
the radiation epoch and for simplicity we consider a loop that was 
formed at the epoch of radiation-matter equality, 
$t_f = t_{\rm eq} \approx 10^{11} {\rm s}$.
With $G\mu \sim 10^{-8}$, 
$t_* \sim t_P/(G\mu)^2 \sim 10^{-27} {\rm s}$ \cite{VilenkinShellard},
where $t_P \approx 10^{-43}$ s is the Planck time, and $\alpha = 0.7$, 
we get
\begin{equation}
\frac{dP_\omega}{dl d\omega} = 
            \frac{(G\mu)^{2+4\alpha}}{t_f^{2-2\alpha} t_P^{2\alpha}}
\approx 10^{-22} ~\frac{\rm ergs}{\rm cm} \ .
\end{equation}
Detectors on Earth observe a flux of photons and it is more relevant
to calculate the number of photons arriving at the detector. This 
follows from $E = N_\omega \omega$ where $N_\omega$ is the number
of photons of frequency $\omega$ emitted by the string, 
\begin{equation}
\frac{dN_\omega}{dt dl} \approx 10^{-22} ~\frac{\rm ergs}{\rm cm} 
                 ~ \frac{d\omega}{\omega} \ .
\label{Ntl}
\end{equation}
If the loop of length $t_{\rm eq} \sim 10^{21} {\rm cm}$ is at a distance 
comparable to the present horizon, $r \sim 10^{27}{\rm cm}$, then the flux 
of photons at 
the detector is obtained by multiplying (\ref{Ntl}) by $t_{\rm eq}/r^2$,
\begin{equation}
\frac{d{\cal N}_\omega}{dt dA} \biggl |_{\rm 1~loop} \approx 
\frac{10^{-10}}{\rm km^2-yr} \frac{d\omega}{\omega} 
\left ( \frac{G\mu}{10^{-8}} \right )^{2(1+2\alpha)}
\label{calN1loop}
\end{equation}
where ${\cal N}_\omega$ denotes the number of photons arriving at the 
detector with collecting area $dA$. If at $t_{\rm eq}$ there was one
loop of length $t_{\rm eq}$ per horizon, the number of loops that can
contribute to the flux at the detector is given by the number of
horizons at $t_{\rm eq}$ that fit within a comoving volume equal to
our present horizon volume: $t_0^3/(t_{\rm eq}z_{\rm eq})^3$, where
$t_0 \sim 10^{17}{\rm s}$ and $z_{\rm eq} \approx (t_0/t_{\rm eq})^{2/3}$.
Therefore the number of contributing loops is $\sim t_0/t_{\rm eq} \sim 10^6$
and the photon flux due to all of these loops is
\begin{equation}
\frac{d{\cal N}_\omega}{dt dA} \biggl |_{\rm loops~at~rec.} \approx 
\frac{10^{-4}}{\rm km^2-yr} \frac{d\omega}{\omega} 
\left ( \frac{G\mu}{10^{-8}} \right )^{2(1+2\alpha)}
\label{calN}
\end{equation}
This estimate suggests that a detector 
with collecting area $~ (100 ~{\rm km})^2$ -- comparable to the Auger 
observatory -- will detect $\sim 1$ photon/year in every logarithmic 
frequency interval emitted by string loops from the recombination era.
The estimate (\ref{calN}) holds for frequencies all the way up to the 
string scale $\sim 10^{15} ~{\rm GeV}$ but it does not take into account 
any propagation effects. Neither does it take into account the network of 
long strings and the spatial and length distribution of loops. Note that the 
emission falls steeply with decreasing string tension. The effect can only 
be useful for small $G\mu$ if the amount of string in a horizon volume is 
inversely proportional to some high power of $G\mu$.

The pattern of photon emission from a string is lineal and this may be helpful
to distinguish it from conventional sources. It is also possible that the 
emission from the string will be polarized (though our analysis so far 
has summed over polarizations and hence erases the polarization information).
The beamed emission from cusps and kinks may provide distinctive events
that can signal the presence of strings. We plan to explore these signatures 
in future work.

We would like to close with a cautionary note. The emission rate
from kink-kink collisions is greatly enhanced by the factor
$(t_f/t_*)^\alpha$ in Eq.~(\ref{avgpsi}). This factor is due to
the accumulation of kinks on strings from the time, $t_*$, when their
dynamics became undamped. The exponent, $\alpha$, depends on dynamical 
factors, such as Hubble expansion, that tend to straighten out the kinks, 
but the estimate does not take radiation backreaction into account and 
this will have a tendency to reduce the emission rate. However, it
is possible that emission at frequencies much lower than the string
scale remain relatively unaffected by the backreaction. We can be
more confident of our estimates of light from cosmic strings only
once this issue is satisfactorily resolved.

\begin{acknowledgments}
We are grateful to Yi-Zen Chu, Ed Copeland and Tom Kibble for comments and 
discussions. TV was supported by the U.S. Department of Energy at Case Western 
Reserve University and also by grant number DE-FG02-90ER40542 at the
Institute for Advanced Study, where this work was partially done. DAS also
thanks the Institute for Advanced Study, where much of this work was done, 
for support.

\end{acknowledgments}



\bibstyle{aps}
\bibliography{paper}

\end{document}